\documentclass[12pt,thmsa]{article}
\usepackage{sw20lart}

\input{tcilatex}
\begin{document}

\title{Baryon Modes of $B$ Meson Decays}
\author{Fayyazuddin \\
%EndAName
National Centre for Physics and Department of Physics\\
Quaid-i-Azam University\\
Islamabad, Pakistan.}
\maketitle

\begin{abstract}
The baryon decay modes of $B,\bar{B}\rightarrow N_{1}\bar{N}_{2}(f)$, $\bar{N%
}_{1}N_{2}(\bar{f})$ provide a frame work to test $CP$-invariance in baryon
sector. It is shown that in the rest frame of $B,$ $N_{1}$ and $\bar{N}_{2}$
come out with longitudnal polarization $\lambda _{1}=\lambda _{2}=\pm 1$
with decay width $\Gamma _{f}=\Gamma _{f}^{++}+\Gamma _{f}^{--}$ and the
asymmetry parameter $\alpha _{f}=\Delta \Gamma _{f}=\Gamma _{f}^{++}-\Gamma
_{f}^{--}$ . It is shown that $CP$ invariance prediction $\alpha _{f}=-\bar{%
\alpha}_{\bar{f}}$ can be tested in these decay modes; especially in the
time dependent decays of $B_{q}^{0}-\bar{B}_{q}^{0}$ complex. Apart from
this, it is shown that decay modes $B(\bar{B})\rightarrow N_{1}\bar{N}_{2}(%
\bar{N}_{1}N_{2})$ and subsequent non leptonic decays of $N_{2},\bar{N}_{2}$
or $(N_{1},\bar{N}_{1})$ into hyperon (antihyperon) also provide a frame
work to study $CP$-odd observables in hyperon decays.
\end{abstract}

\section{Introduction}

The $CP$-violation in kaon and $B_{q}^{0}-\bar{B}_{q}^{0}$ systems has been
extensively studied \cite{1}. There is thus a need to study $CP$-violation
outside these systems. In hyperon decays, the observables are the decay rate 
$\Gamma ,$ asymmetry parameter $\alpha ,$ the transverse polarization $\beta 
$ and longitudnal polarization $\gamma $ \cite{2}. $CP$ asymmetry predicts $%
\bar{\Gamma}=\Gamma ,$ $\bar{\alpha}=-\alpha ,$ $\bar{\beta}=-\beta ,$ where
these observables correspond to non-leptonic hyperon decays $N\rightarrow
N^{\prime }\pi $ and $\bar{N}\rightarrow \bar{N}^{\prime }\bar{\pi}$. Thus
to leading order $CP$ -odd observables are \cite{3} 
\begin{equation}
\delta \Gamma =\frac{\Gamma -\bar{\Gamma}}{\Gamma +\bar{\Gamma}},\text{ }%
\delta \alpha =\frac{\alpha +\bar{\alpha}}{\alpha -\bar{\alpha}},\text{ }%
\delta \beta =\frac{\beta +\bar{\beta}}{\beta -\bar{\beta}}  \label{1}
\end{equation}
`The decays of $B(\bar{B})$ mesons to baryon-antibaryon pair $N_{1}$ $\bar{N}%
_{2}$ $(\bar{N}_{1}$ $N_{2})$ and subsequent decays of $N_{2},\bar{N}_{2}$
or $(N_{1},\bar{N}_{1})$ to a lighter hyperon (antihyperon) plus a meson
provide a means to study $CP$-odd observables as for example in the process 
\[
e^{-}e^{+}\rightarrow B,\bar{B}\rightarrow N_{1}\bar{N}_{2}\rightarrow N_{1}%
\bar{N}_{2}^{\prime }\bar{\pi},\text{ }\bar{N}_{1}N_{2}\rightarrow \bar{N}%
_{1}N_{2}^{\prime }\pi 
\]
Apart from the above motivation, the baryon decay modes of $B$-mesons are of
intrinsic intrest by themselves as we discuss below. The baryon decay modes
of $B_{d}^{0}-\bar{B}_{d}^{0}$ have also been discussed in a differnt
context in \cite{4}.

In the rest frame of $B$, $N_{1}$ and $\bar{N}_{2}$ come out longitudnally
polarized with polarization 
\[
\left( \lambda _{1}\equiv \frac{E_{1}}{m_{1}}\mathbf{n}\cdot \mathbf{s}%
_{1}\right) =\left( \lambda _{2}\equiv \frac{E_{2}}{m_{2}}(-\mathbf{n}\cdot 
\mathbf{s}_{2})\right) =\pm 1, 
\]
where 
\begin{eqnarray*}
\mathbf{p}_{1} &=&\left| \mathbf{p}\right| \mathbf{n},\mathbf{p}_{2}=-\left| 
\mathbf{p}\right| \mathbf{n},\mathbf{s}_{1}=\frac{m_{1}}{E_{1}}\mathbf{n} \\
\mathbf{s}_{2} &=&-\frac{m_{2}}{E_{2}}\mathbf{n}
\end{eqnarray*}
$s_{1}^{\mu },s_{2}^{\mu }$ are polarization vectors of $N_{1}$ and $\bar{N}%
_{2}$ respectively $(p_{1}\cdot s_{1}=0,$ $p_{2}\cdot s_{2}=0,$ $%
s_{1}^{2}=-1=s_{2}^{2}).$ The decay $B\rightarrow N_{1}\bar{N}_{2}(f)$ is
described by the matrix element 
\begin{equation}
M_{f}=F_{q}e^{+i\phi }\left[ \bar{u}(\mathbf{p}_{1})(A_{f}+\gamma
_{5}B_{f})v(\mathbf{p}_{2})\right]  \label{2}
\end{equation}
where $F_{q}$ is a constant containing CKM factor, $\phi $ is the weak
phase. The amplitude $A_{f}$ and $B_{f}$ are in general complex in the sense
that they incorporate the final state phases $\delta _{p}^{f}$ and $\delta
_{s}^{f}.$ Note that $A_{f}$ is the parity violating amplitude ($p$-wave)
whereas $B_{f}$ is parity conserving amplitude ($s$-wave). The $CPT$
invariance gives the matrix elements for the decay $\bar{B}\rightarrow \bar{N%
}_{1}N_{2}(\bar{f}):$%
\begin{equation}
\bar{M}_{\bar{f}}=F_{q}e^{-i\phi }\left[ \bar{u}(\mathbf{p}%
_{2})(-A_{f}+\gamma _{5}B_{f})v(\mathbf{p}_{1})\right]  \label{3}
\end{equation}
If the decays are described by a single matrix element $M_{f},$ then $CPT$
and $CP$ invariance give the same prediction viz 
\begin{equation}
\bar{\Gamma}_{\bar{f}}=\Gamma _{f},\text{ }\bar{\alpha}_{\bar{f}}=-\alpha
_{f},\text{ }\bar{\beta}_{\bar{f}}=-\beta _{f},\text{ }\bar{\gamma}_{\bar{f}%
}=\gamma _{f}  \label{4}
\end{equation}

\section{Decay Rate and Asymmetry Parameters:}

The decay width for the mode $B\rightarrow N_{1}\bar{N}_{2}(f)$ is given by 
\begin{eqnarray}
\Gamma _{f} &=&\frac{m_{1}m_{2}}{2\pi m_{B}^{2}}\left| \mathbf{p}\right|
\left| M_{f}\right| ^{2}  \nonumber \\
&=&\frac{F_{q}^{2}}{2\pi m_{B}^{2}}\left| \mathbf{p}\right| \left[
(p_{1}\cdot p_{2}-m_{1}m_{2})\left| A_{f}\right| ^{2}+(p_{1}\cdot
p_{2}+m_{1}m_{2})\left| B_{f}\right| ^{2}\right]  \label{5}
\end{eqnarray}
In order to take into account the polarization of $N_{1}$ and $\bar{N}_{2},$
we give the general expression for $\left| M_{f}\right| ^{2}$%
\begin{eqnarray}
\left| M_{f}\right| ^{2} &=&\frac{F_{q}^{2}}{16m_{1}m_{2}}Tr\left[ 
\begin{array}{c}
(\not{p}_{1}+m_{1})(1+\gamma _{5}\gamma \cdot s_{1})(A_{f}+\gamma _{5}B_{f})(%
\not{p}_{2}-m_{2}) \\ 
\times (1+\gamma _{5}\gamma \cdot s_{2})(A_{f}^{*}-\gamma _{5}B_{f}^{*})
\end{array}
\right]  \nonumber \\
&=&\frac{4F_{q}^{2}}{16m_{1}m_{2}}\left[ 
\begin{array}{c}
\left| A_{f}\right| ^{2}(p_{1}\cdot p_{2}-m_{1}m_{2})+\left| B_{f}\right|
^{2}(p_{1}\cdot p_{2}+m_{1}m_{2}) \\ 
-(A_{f}B_{f}^{*}+B_{f}A_{f}^{*})(m_{2}p_{1}\cdot s_{2}+m_{1}p_{2}\cdot s_{1})
\\ 
-i(A_{f}B_{f}^{*}-B_{f}A_{f}^{*})(\epsilon ^{\mu \nu \rho \lambda
}p_{1}^{\mu }s_{1}^{\nu }p_{2}^{\rho }s_{2}^{\lambda }) \\ 
+m_{1}m_{2}(\left| A_{f}\right| ^{2}+\left| B_{f}\right| ^{2})s_{1}\cdot
s_{2} \\ 
+\left( \left| A_{f}\right| ^{2}-\left| B_{f}\right| ^{2}\right) \left(
-p_{1}\cdot p_{2}s_{1}\cdot s_{2}+(p_{1}\cdot s_{2})(p_{2}\cdot s_{1})\right)
\end{array}
\right]  \label{6}
\end{eqnarray}
It is clear that Eqs. (\ref{4}) follows from Eqs.(\ref{2}) and (\ref{6}). In
the rest frame of $B$, we get from Eqs.(\ref{5}) and (\ref{6}) 
\begin{equation}
\left| M_{f}\right| ^{2}=F_{q}^{2}\frac{2E_{1}E_{2}}{4m_{1}m_{2}}\left[
\left| a_{s}^{f}\right| ^{2}+\left| a_{p}^{f}\right| ^{2}\right] \left\{ 
\begin{array}{c}
1+\alpha _{f}\left( \frac{m_{1}}{E_{1}}\mathbf{n}_{1}\cdot \mathbf{s}_{1}-%
\frac{m_{2}}{E_{2}}\mathbf{n}\cdot \mathbf{s}_{2}\right) \\ 
+\beta _{f}\mathbf{n}\cdot (\mathbf{s}_{1}\times \mathbf{s}_{2})+\gamma
_{f}\left[ (\mathbf{n}_{1}\cdot \mathbf{s}_{1})(\mathbf{n}\cdot \mathbf{s}%
_{2})-\mathbf{s}_{1}\cdot \mathbf{s}_{2}\right] \\ 
-\frac{m_{1}m_{2}}{E_{1}E_{2}}(\mathbf{n}\cdot \mathbf{s}_{1})(\mathbf{n}%
\cdot \mathbf{s}_{2})
\end{array}
\right\}  \label{7}
\end{equation}
where 
\begin{eqnarray}
a_{s} &=&\sqrt{\frac{p_{1}\cdot p_{2}+m_{1}m_{2}}{2E_{1}E_{2}}}B,\text{ }%
a_{p}=-\sqrt{\frac{p_{1}\cdot p_{2}-m_{1}m_{2}}{2E_{1}E_{2}}}A  \label{8} \\
\alpha _{f} &=&\frac{2S_{f}P_{f}\cos (\delta _{s}^{f}-\delta _{p}^{f})}{%
S_{f}^{2}+P_{f}^{2}},\text{ }\beta _{f}=\frac{2S_{f}P_{f}\sin (\delta
_{s}^{f}-\delta _{p}^{f})}{S_{f}^{2}+P_{f}^{2}}  \nonumber \\
\gamma _{f} &=&\frac{S_{f}^{2}-P_{f}^{2}}{S_{f}^{2}+P_{f}^{2}},\text{ }%
a_{s}=S_{f}e^{i\delta _{s}^{f}},a_{p}^{f}=P_{f}e^{i\delta _{p}^{f}}
\label{9}
\end{eqnarray}
However in the rest frame of $B$, due to spin conservation 
\begin{equation}
\frac{E_{1}}{m_{1}}\mathbf{n}\cdot \mathbf{s}_{1}=\frac{E_{2}}{m_{2}}(-%
\mathbf{n}\cdot \mathbf{s}_{2})=\pm 1  \label{10}
\end{equation}
Thus invariants multiplying $\beta _{f}$ and $\gamma _{f}$ vanish. Hence we
have 
\begin{eqnarray}
\left| M_{f}\right| ^{2} &=&\left( \frac{2E_{1}E_{2}}{m_{1}m_{2}}\right)
F_{q}^{2}(S_{f}^{2}+P_{f}^{2})\left[ (1+\lambda _{1}\lambda _{2})+\alpha
_{f}(\lambda _{1}+\lambda _{2})\right]  \label{11} \\
\Gamma _{f} &=&\Gamma _{f}^{++}+\Gamma _{f}^{--}=\frac{2E_{1}E_{2}}{2\pi
m_{B}^{2}}\left| \vec{p}\right| F_{q}^{2}\left[ S_{f}^{2}+P_{f}^{2}\right] =%
\bar{\Gamma}_{\bar{f}}  \label{12} \\
\Delta \Gamma _{f} &=&\frac{\Gamma _{f}^{++}-\Gamma _{f}^{--}}{\Gamma
_{f}^{++}+\Gamma _{f}^{--}}=\alpha _{f}\text{ ; }\Delta \bar{\Gamma}_{\bar{f}%
}=\bar{\alpha}_{\bar{f}}=-\alpha _{f}  \label{13}
\end{eqnarray}
Eqs.(\ref{12}) and (\ref{13}) follow from $CP$ or $CPT$ invariance. It will
be of intrest to test these equations.\thinspace

In this paper, we confine ourself to decays $B\rightarrow N_{1}\bar{N}_{2}(%
\bar{B}\rightarrow \bar{N}_{1}N_{2})$ described by a single matrix element $%
M_{f}$ $(\bar{M}_{\bar{f}})$ i.e. to the effective Lagrangians 
\begin{eqnarray}
\mathcal{L} &=&V_{cb}V_{uq}^{*}[\bar{q}u]_{V-A}[\bar{c}b]_{V-A}+h.c.
\label{14} \\
\mathcal{L} &=&V_{ub}V_{cq}^{*}[\bar{q}c]_{V-A}[\bar{u}b]_{V-A}+h.c
\label{15}
\end{eqnarray}
where $q=d$ or $s.$ For the decay modes described by the above Lagrangians,
there are no contributions from the penguin diagrams. The Lagrangian given
in Eq.(\ref{14}) is relevant for the decays 
\begin{eqnarray*}
\text{i) }B_{q}^{0} &\rightarrow &N_{1}\bar{N}_{2}(f);\text{ }\bar{B}%
_{q}^{0}\rightarrow \bar{N}_{1}N_{2}(\bar{f}) \\
N_{1}N_{2} &:&p\Lambda _{c}^{+},\Sigma ^{+}\Xi _{c}^{+},\frac{1}{\sqrt{6}}%
\Lambda \Xi _{c}^{0},\frac{1}{\sqrt{2}}\Sigma ^{0}\Xi _{c}^{0} \\
B_{c}^{+} &\rightarrow &p\bar{n},\Sigma ^{+}\bar{\Lambda}(q=d);\text{ }%
B_{c}^{+}\rightarrow p\bar{\Lambda},\text{ }\Sigma ^{+}\bar{\Xi}^{0}\text{ }%
(q=s)
\end{eqnarray*}
For the decay modes(i), the weak phase $\phi =0$ and the decay matrix
elements $M_{f}$ and $\bar{M}_{\bar{f}}$ are given by Eqs.(\ref{2}) and (\ref
{3}). For the Lagrangian given in Eq.(\ref{15}), the relevant decay modes
are 
\begin{eqnarray*}
\text{ii) }\bar{B}_{q}^{0} &\rightarrow &N_{1}\bar{N}_{2}(f);\text{ }%
B_{q}^{0}\rightarrow \bar{N}_{1}N_{2}(\bar{f}) \\
B^{-} &\rightarrow &N_{1}\bar{N}_{2}:\text{ }n\bar{\Lambda}_{c}^{-},\text{ }%
\frac{1}{\sqrt{6}}\Lambda \bar{\Xi}_{c}^{-},\text{ }-\frac{1}{\sqrt{2}}%
\Sigma ^{0}\bar{\Xi}_{c}^{-},\text{ }\Sigma ^{-}\bar{\Xi}_{c}^{0}\text{ }%
(q=d)
\end{eqnarray*}
\[
-\frac{2}{\sqrt{6}}\Lambda \bar{\Lambda}_{c}^{-},\text{ }\Xi ^{0}\bar{\Xi}%
_{c}^{-},\text{ }\Xi ^{-}\bar{\Xi}_{c}^{0}\text{ }(q=s) 
\]
For various decay channels (i) and (ii), we have explicitly shown the $SU(3)$
factors. For the decay modes (ii), the weak phase $\phi =\phi _{3}/\gamma ,$
which arises from $V_{ub}=\left| V_{ub}\right| e^{-i\gamma }$. For the decay
modes (ii), the matrix elements $\bar{M}_{f}^{\prime }$ and $M_{\bar{f}%
}^{\prime }$ are given by 
\begin{eqnarray}
\bar{M}_{f}^{\prime } &=&e^{-i\phi _{3}}F_{q}^{\prime }[\bar{u}(\mathbf{p}%
_{1})(A_{\bar{f}}^{\prime }+\gamma _{5}B_{\bar{f}}^{\prime })v(\mathbf{p}%
_{2})]  \label{16} \\
M_{\bar{f}}^{\prime } &=&e^{i\phi _{3}}F_{q}^{\prime }[\bar{u}(\mathbf{p}%
_{2})(-A_{\bar{f}}^{\prime }+\gamma _{5}B_{\bar{f}}^{\prime })v(\mathbf{p}%
_{1})]  \label{17}
\end{eqnarray}

Hence the decay widths and $CP-$asymmetry parameters are given by 
\begin{eqnarray}
\bar{\Gamma}_{f}^{\prime } &=&\Gamma _{\bar{f}}^{\prime }=\frac{2E_{1}E_{2}}{%
8\pi m_{B}^{2}}\left| \mathbf{p}\right| F_{q}^{\prime 2}(S_{\bar{f}}^{\prime
2}+P_{\bar{f}}^{\prime 2})  \label{18} \\
\bar{\alpha}_{f}^{\prime } &=&-\alpha _{\bar{f}}^{\prime }=\frac{2S_{\bar{f}%
}^{\prime }P_{\bar{f}}^{\prime }\cos (\delta _{s}^{\bar{f}}-\delta _{p}^{%
\bar{f}})}{(S_{\bar{f}}^{\prime 2}+P_{\bar{f}}^{\prime 2})}  \label{19}
\end{eqnarray}
Now 
\begin{eqnarray}
F_{q} &=&\frac{G_{F}}{\sqrt{2}}(a_{2},a_{1})V_{cb}V_{uq}  \label{20} \\
F_{q}^{\prime } &=&\frac{G_{F}}{\sqrt{2}}(a_{2},a_{1})\left| V_{ub}\right|
V_{cq}  \label{21}
\end{eqnarray}
Define 
\begin{eqnarray}
r &=&\frac{F_{q}^{\prime }}{F_{q}}=\frac{\left| V_{ub}\right| V_{cq}}{%
V_{cb}V_{uq}}=-\lambda ^{2}\sqrt{\bar{\rho}^{2}+\bar{\eta}^{2}}\text{ for }%
q=d  \label{22} \\
&=&\sqrt{\bar{\rho}^{2}+\bar{\eta}^{2}}\text{ for }q=s  \nonumber
\end{eqnarray}
$a_{2}$ $(a_{1})$ are factors which account for color supressed (without
color supressed) matrix elements. From Eqs.(\ref{12}), (\ref{20}), we get 
\begin{eqnarray}
\frac{\Gamma (B_{s}^{0}\rightarrow p\bar{\Lambda}_{c}^{-})}{\Gamma
(B_{d}^{0}\rightarrow p\bar{\Lambda}_{c}^{-})} &=&\lambda ^{2}\left( \frac{%
m_{B_{d}}}{m_{B_{s}}}\right) ^{2}\frac{\left[ E_{1}E_{2}\left| \vec{p}%
\right| \right] _{B_{s}}}{\left[ E_{1}E_{2}\left| \vec{p}\right| \right]
_{B_{d}}}\xi ^{2}  \nonumber \\
&\approx &\lambda ^{2}\xi ^{2}  \label{23}
\end{eqnarray}
where $\xi $ is a measure of $SU(3)$ violation.

Now $B_{q}^{0},$ $\bar{B}_{q}^{0}$ annihilate into baryon-antibaryon pair $%
N_{1}\bar{N}_{2}$ through $W$-exchange as depicted in Figs (1a) and (1b). $%
B^{-}\rightarrow N_{1}\bar{N}_{2}$ through annihilation diagram is shown in
Fig (2). It is clear from Fig (1a) and (1b), that we have the same final
state configuration for $B_{q}^{0},$ $\bar{B}_{q}^{0}\rightarrow N_{1}\bar{N}%
_{2}.$ Thus one would expect 
\begin{eqnarray}
S_{\bar{f}}^{\prime } &=&S_{f},\text{ }P_{\bar{f}}^{\prime }=P_{f}  \nonumber
\\
\delta _{s}^{\bar{f}} &=&\delta _{s}^{f},\delta _{p}^{\bar{f}}=\delta
_{p}^{f}  \label{24}
\end{eqnarray}
Hence we have 
\begin{eqnarray}
\Gamma _{\bar{f}}^{\prime } &=&\bar{\Gamma}_{f}^{\prime }=r^{2}\Gamma _{f}
\label{25} \\
\bar{\alpha}_{f}^{\prime } &=&-\alpha _{\bar{f}}^{\prime }=\alpha _{f}=-\bar{%
\alpha}_{\bar{f}}  \label{26} \\
\frac{\Gamma (\bar{B}_{s}^{0}\rightarrow p\bar{\Lambda}_{c}^{-})}{\Gamma
(B_{s}^{0}\rightarrow p\bar{\Lambda}_{c}^{-})} &=&\left( \bar{\rho}^{2}+\bar{%
\eta}^{2}\right)  \label{27} \\
\frac{\Gamma (B^{-}\rightarrow \Lambda \bar{\Lambda}_{c}^{-})}{\Gamma
(B_{d}^{0}\rightarrow p\bar{\Lambda}_{c}^{-})} &\approx &\frac{2}{3}\left(
\lambda \frac{a_{1}}{a_{2}}\right) ^{2}\left( \bar{\rho}^{2}+\bar{\eta}%
^{2}\right)  \label{28}
\end{eqnarray}
Eq.(\ref{28}) is valid in $SU(3)$ limit, but $SU(3)$ breaking effects can be
taken into account by using physical masses for proton and $\Lambda $
hyperon in the kinematical factors.

Above predictions can be tested in future experiments on baryon decay modes
of $B$-mesons. In particular $\bar{\alpha}_{f}^{\prime }=\alpha _{f}$ would
give direct confirmation of Eqs.(\ref{24}).

Finally, we discuss $B_{d}^{0}\rightarrow p\bar{\Lambda}_{c}^{-}$ decay. For
this decay mode the experimental branching ratio is $(2.2\pm 0.8)\times
10^{-5}$\cite{5}. Using the experimental value for $\tau _{B_{d}^{0},}$ we
obtain 
\begin{equation}
\Gamma (B_{d}^{0}\rightarrow p\bar{\Lambda}_{c}^{-})=(9.46\pm 3.44)\times
10^{-15}\text{ MeV}  \label{29}
\end{equation}
The decay width in terms of $[S_{f}^{2}+P_{f}^{2}]$ is given by 
\begin{equation}
\Gamma _{f}=\frac{G_{F}^{2}}{2}\left| V_{cb}\right| ^{2}\left| V_{ud}\right|
^{2}a_{2}^{2}\left( S_{f}^{2}+P_{f}^{2}\right) \left[ \frac{2E_{1}E_{2}}{%
2\pi m_{B}^{2}}\left| \mathbf{p}\right| \right]  \label{30}
\end{equation}
Using $\left| V_{cb}\right| =41.6\times 10^{-3},$ $\left| V_{ud}\right|
=0.97378$ \cite{5}, $a_{2}=0.226$ and noting that 
\[
\frac{2E_{1}E_{2}\left| \mathbf{p}\right| }{2m_{B}^{2}}\approx 1.01\text{ GeV%
} 
\]
we get 
\begin{equation}
\Gamma _{f}=[9.09\times 10^{-25}\text{MeV}^{-3}][S_{f}^{2}+P_{f}^{2}]
\label{31}
\end{equation}
Using Eq.(\ref{29}), we get 
\begin{equation}
\left( S_{f}^{2}+P_{f}^{2}\right) =(1.04\pm 0.38)\times 10^{10}\text{MeV}^{4}
\label{32}
\end{equation}
In order to express $\left( S_{f}^{2}+P_{f}^{2}\right) $ in terms of
dimensionless form factors, we use $B^{-}\rightarrow l^{-}\bar{\nu}_{l}$
decay as a guide, which also occurs through a diagram similar to Fig 2.

For the decay $B^{-}\rightarrow l^{-}\nu _{l},$%
\begin{eqnarray}
\Gamma (B^{-} &\rightarrow &l^{-}\bar{\nu}_{l})=\frac{G_{F}^{2}}{2}\left|
V_{ub}\right| ^{2}\left( \frac{2E_{1}E_{2}}{2\pi m_{B}^{2}}\right) \left| 
\mathbf{p}\right| \left[ S^{2}+P^{2}\right]  \nonumber \\
&=&\frac{G_{F}^{2}}{2}\left| V_{ub}\right| ^{2}\left( \frac{2E_{1}E_{2}}{%
2\pi m_{B}^{2}}\right) \left| \mathbf{p}\right| 2(m_{l}^{2}+m_{\nu
_{l}}^{2})f_{B}^{2}  \label{33}
\end{eqnarray}
Noting that 
\[
\frac{2E_{1}E_{2}\left| \mathbf{p}\right| }{m_{B}^{2}}\approx \frac{1}{4}%
m_{B} 
\]
we get 
\begin{equation}
\Gamma (B^{-}\rightarrow l^{-}\bar{\nu}_{l})\approx \frac{G_{F}^{2}}{8\pi }%
\left| V_{ub}\right| ^{2}m_{B}m_{l}^{2}f_{B}^{2}  \label{34}
\end{equation}
Thus we see that for this decay 
\begin{equation}
S^{2}+P^{2}=2(m_{l}^{2}+m_{\nu _{l}}^{2})f_{B}^{2}  \label{35}
\end{equation}
Hence we can parametrize $(S_{f}^{2}+P_{f}^{2})$ in terms of two form
factors $F_{V}^{\Lambda _{c}-p}(s)$ and $F_{A}^{\Lambda _{c}-p}(s):$%
\begin{eqnarray}
P_{f}^{2} &=&f_{B}^{2}(m_{\Lambda _{c}}+m_{p})^{2}\left[ \left( \frac{%
m_{\Lambda _{c}}-m_{p}}{m_{\Lambda _{c}}+m_{p}}\right) F_{V}^{\Lambda
_{c}-p}(s)\right] _{s=m_{B}^{2}}^{2}  \nonumber \\
S_{f}^{2} &=&f_{B}^{2}(m_{\Lambda _{c}}+m_{p})^{2}\left[ F_{A}^{\Lambda
_{c}-p}(s)\right] _{s=m_{B}^{2}}^{2}  \label{36}
\end{eqnarray}
It is easy to see that for $F_{V}=1$ and $F_{A}=1,$ it reduces to form of
Eq.(\ref{35}). Using the experimental values for the masses and $%
f_{B}\approx 180$ MeV, we get from Eq.(\ref{33}) 
\begin{equation}
(0.175)[F_{V}^{\Lambda _{c}-p}(m_{B}^{2})]^{2}+[F_{A}^{\Lambda
_{c}-p}(m_{B}^{2})]^{2}=(3.1\pm 1.1)\times 10^{-2}  \label{37}
\end{equation}
The dominant contribution comes from the axial vector form factor. The decay 
$B_{c}^{-}\rightarrow n\bar{p}$ would give information for nucleon form
factors: 
\begin{eqnarray}
P_{f}^{2} &=&f_{B_{c}}^{2}(m_{n}+m_{p})^{2}\left[ \frac{m_{n}-m_{p}}{%
m_{n}+m_{p}}F_{V}(s)\right] _{s=m_{B_{c}}^{2}}^{2}\approx 0  \nonumber \\
S_{f}^{2} &=&f_{B_{c}}^{2}(m_{n}+m_{p})^{2}\left[ F_{A}^{2}(s)\right]
_{s=m_{B_{c}}^{2}}  \label{38}
\end{eqnarray}
The baryon decay modes of $B$ -mesons also provide the means to explore the
baryon form factors at high $s.$ Finally, we note that Eq.(\ref{36}), give
the $SU(3)$ breaking factor $\xi =\frac{f_{B_{s}}}{f_{B}}$ in Eq.(\ref{23}).

\section{Time- Dependent Baryon Decay Modes of $B_{q}^{0}$}

Define the amplitudes 
\begin{eqnarray}
\mathcal{A}^{\lambda _{1}\lambda _{2}}(t) &=&\frac{\left[ \Gamma
(B_{q}^{0}(t)\rightarrow f)-\Gamma (\bar{B}_{q}^{0}(t)\rightarrow \bar{f}%
)\right] _{\lambda _{1}\lambda _{2}}+\left[ \Gamma (B_{q}^{0}(t)\rightarrow 
\bar{f})-\Gamma (\bar{B}_{q}^{0}(t)\rightarrow f)\right] _{\lambda
_{1}\lambda _{2}}}{\sum_{\lambda _{1}\lambda _{2}}\left[ \Gamma
(B_{q}^{0}(t)\rightarrow f,\bar{f})+\Gamma (\bar{B}_{q}^{0}(t)\rightarrow 
\bar{f},f)\right] _{\lambda _{1}\lambda _{2}}}  \nonumber \\
&=&\frac{-2\sin \Delta mt\left[ \func{Im}e^{2i\phi _{M}}(M_{f}^{*}\bar{M}%
_{f}^{\prime }+M_{f}^{\prime *}\bar{M}_{\bar{f}})\right] }{\sum_{\lambda
_{1}\lambda _{2}}\left[ \left| M_{f}^{2}\right| +\left| \bar{M}_{\bar{f}%
}^{2}\right| +\left| M_{\bar{f}}^{\prime 2}\right| +\left| M_{f}^{\prime
2}\right| \right] }  \nonumber \\
&&  \label{39a} \\
\mathcal{F}^{\lambda _{1}\lambda _{2}}(t) &=&\frac{\left[ \Gamma
(B_{q}^{0}(t)\rightarrow f)-\Gamma (\bar{B}_{q}^{0}(t)\rightarrow \bar{f}%
)\right] _{\lambda _{1}\lambda _{2}}-\left[ \Gamma (B_{q}^{0}(t)\rightarrow 
\bar{f})-\Gamma (\bar{B}_{q}^{0}(t)\rightarrow f)\right] _{\lambda
_{1}\lambda _{2}}}{\sum_{\lambda _{1}\lambda _{2}}\left[ \Gamma
(B_{q}^{0}(t)\rightarrow f,\bar{f})+\Gamma (\bar{B}_{q}^{0}(t)\rightarrow 
\bar{f},f)\right] _{\lambda _{1}\lambda _{2}}}  \nonumber \\
&=&\frac{\cos \Delta mt\left[ \left| M_{f}^{2}\right| +\left| \bar{M}_{\bar{f%
}}^{2}\right| -\left| M_{\bar{f}}^{\prime 2}\right| -\left| \bar{M}%
_{f}^{\prime 2}\right| \right] -2\sin \Delta mt\left[ \func{Im}e^{2i\phi
_{M}}(M_{f}^{*}\bar{M}_{f}^{\prime }-M_{f}^{\prime *}\bar{M}_{\bar{f}%
})\right] }{\sum_{\lambda _{1}\lambda _{2}}\left[ \left| M_{f}^{2}\right|
+\left| \bar{M}_{\bar{f}}^{2}\right| +\left| M_{\bar{f}}^{\prime 2}\right|
+\left| M_{f}^{\prime 2}\right| \right] }  \nonumber \\
&&  \label{39}
\end{eqnarray}
Thus 
\begin{eqnarray}
&&8\left[ (S_{f}^{2}+P_{f}^{2})+r^{2}(S_{\bar{f}}^{\prime 2}+P_{\bar{f}%
}^{\prime 2})\right] \mathcal{A}^{\lambda _{1}\lambda _{2}}(t)  \nonumber \\
&=&2\sin \Delta mt\left\{ 
\begin{array}{c}
\sin (2\phi _{M}-\gamma )\left[ 2r(1+\lambda _{1}\lambda _{2})(S_{f}S_{\bar{f%
}}^{\prime }\cos (\delta _{s}^{f}-\delta _{s}^{\bar{f}})+P_{f}P_{\bar{f}%
}^{\prime }\cos (\delta _{p}^{f}-\delta _{p}^{\bar{f}})\right] \\ 
-\cos (2\phi _{M}-\gamma )\left[ 2r(\lambda _{1}+\lambda _{2})(S_{f}P_{\bar{f%
}}^{\prime }\sin (\delta _{s}^{f}-\delta _{p}^{\bar{f}})+S_{\bar{f}}^{\prime
}P_{f}\sin (\delta _{p}^{f}-\delta _{s}^{\bar{f}})\right]
\end{array}
\right\}  \nonumber \\
&&  \label{40}
\end{eqnarray}
\begin{eqnarray}
&&\left. 8\left[ 
\begin{array}{c}
(S_{f}^{2}+P_{f}^{2}) \\ 
+r^{2}(S_{\bar{f}}^{\prime 2}+P_{\bar{f}}^{\prime 2})
\end{array}
\right] \mathcal{F}^{\lambda _{1}\lambda _{2}}(t)\right. =\left. 
\begin{array}{c}
\cos \Delta mt\left\{ (S_{f}^{2}+P_{f}^{2})\left. 
\begin{array}{c}
\left[ 
\begin{array}{c}
2(1+\lambda _{1}\lambda _{2}) \\ 
+(\alpha _{f}+\bar{\alpha}_{\bar{f}})(\lambda _{1}+\lambda _{2})
\end{array}
\right] \\ 
-r^{2}(S_{\bar{f}}^{\prime 2}+P_{\bar{f}}^{\prime 2}) \\ 
\left[ 
\begin{array}{c}
2(1+\lambda _{1}\lambda _{2}) \\ 
+(\bar{\alpha}_{f}^{\prime }+\alpha _{\bar{f}}^{\prime })(\lambda
_{1}+\lambda _{2})
\end{array}
\right]
\end{array}
\right. \right\} \\ 
-2\sin \Delta mt\left\{ 
\begin{array}{c}
\cos (2\phi _{M}-\gamma )\left[ 
\begin{array}{c}
-2r(1+\lambda _{1}\lambda _{2}) \\ 
(S_{f}S_{\bar{f}}^{\prime }\sin (\delta _{s}^{f}-\delta _{s}^{\bar{f}}) \\ 
+P_{f}P_{\bar{f}}^{\prime }\sin (\delta _{p}^{f}-\delta _{p}^{\bar{f}}))
\end{array}
\right] \\ 
+\sin (2\phi _{M}-\gamma )\left[ 
\begin{array}{c}
2r(\lambda _{1}+\lambda _{2}) \\ 
(S_{f}P\cos (\delta _{s}^{f}-\delta _{p}^{\bar{f}}) \\ 
+P_{f}S_{\bar{f}}^{\prime }\cos (\delta _{p}^{f}-\delta _{s}^{\bar{f}})
\end{array}
\right]
\end{array}
\right\}
\end{array}
\right.  \nonumber \\
&&  \label{41}
\end{eqnarray}
These are general expressions for the time-dependent decay modes in the rest
frame of $B_{q}^{0}.$ From Eqs.(\ref{40}) and (\ref{41}), the even and odd
time-dependent decay amplitudes are given by 
\begin{eqnarray}
\mathcal{A(}t) &\equiv &(\mathcal{A}^{++}\mathcal{(}t)+\mathcal{A}^{--}%
\mathcal{(}t))  \nonumber \\
&=&\frac{2r\sin \Delta mt\sin (2\phi _{M}-\gamma )\left[ S_{f}S_{\bar{f}%
}^{\prime }\cos (\delta _{s}^{f}-\delta _{s}^{\bar{f}})+P_{f}P_{\bar{f}%
}^{\prime }\cos (\delta _{p}^{f}-\delta _{p}^{\bar{f}})\right] }{%
(S_{f}^{2}+P_{f}^{2})+r^{2}(S_{\bar{f}}^{\prime 2}+P_{\bar{f}}^{\prime 2})} 
\nonumber \\
&&  \label{42} \\
\Delta \mathcal{A(}t) &\equiv &\mathcal{A}^{++}\mathcal{(}t)-\mathcal{A}^{--}%
\mathcal{(}t)  \nonumber \\
&=&\frac{-2r\sin \Delta mt\cos (2\phi _{M}-\gamma )\left[ S_{f}P_{\bar{f}%
}^{\prime }\sin (\delta _{s}^{f}-\delta _{p}^{\bar{f}})+S_{\bar{f}}^{\prime
}P_{f}\cos (\delta _{p}^{f}-\delta _{s}^{\bar{f}})\right] }{%
(S_{f}^{2}+P_{f}^{2})+r^{2}(S_{\bar{f}}^{\prime 2}+P_{\bar{f}}^{\prime 2})} 
\nonumber \\
&&  \label{43} \\
\mathcal{F}(t) &=&\mathcal{F}^{++}(t)+\mathcal{F}^{--}(t)  \nonumber \\
&=&\frac{%
\begin{array}{c}
\cos \Delta mt\left[ (S_{f}^{2}+P_{f}^{2})-r^{2}(S_{\bar{f}}^{\prime 2}+P_{%
\bar{f}}^{\prime 2})\right] +2r\sin \Delta mt\cos (2\phi _{M}-\gamma ) \\ 
\times \left[ S_{f}S_{\bar{f}}^{\prime }\sin (\delta _{s}^{f}-\delta _{s}^{%
\bar{f}})+P_{f}P_{f}^{\prime }\sin (\delta _{p}^{f}-\delta _{p}^{\bar{f}%
})\right]
\end{array}
}{2\left[ (S_{f}^{2}+P_{f}^{2})+r^{2}(S_{\bar{f}}^{\prime 2}+P_{\bar{f}%
}^{\prime 2})\right] }  \nonumber \\
&&  \label{44} \\
\Delta \mathcal{F}(t) &\equiv &\mathcal{F}^{++}(t)-\mathcal{F}^{--}(t) 
\nonumber \\
&=&\frac{\cos \Delta mt\left[ (S_{f}^{2}+P_{f}^{2})(\alpha _{f}+\bar{\alpha}%
_{\bar{f}})-r^{2}(S_{\bar{f}}^{\prime 2}+P_{\bar{f}}^{\prime 2})(\bar{\alpha}%
_{f}^{\prime }+\alpha _{\bar{f}}^{\prime })\right] }{2\left[
(S_{f}^{2}+P_{f}^{2})+r^{2}(S_{\bar{f}}^{\prime 2}+P_{\bar{f}}^{\prime
2})\right] }  \nonumber \\
&&-\frac{2r\sin \Delta mt\sin (2\phi _{M}-\gamma )\left[ S_{f}P_{\bar{f}%
}^{\prime }\cos (\delta _{s}^{f}-\delta _{p}^{\bar{f}})+P_{f}S_{\bar{f}%
}^{\prime }\cos (\delta _{p}^{f}-\delta _{s}^{\bar{f}})\right] }{%
(S_{f}^{2}+P_{f}^{2})+r^{2}(S_{\bar{f}}^{\prime 2}+P_{\bar{f}}^{\prime 2})} 
\nonumber \\
&&  \label{45}
\end{eqnarray}
For $B_{d}^{0},$ $r=-\lambda ^{2}\sqrt{\bar{\rho}^{2}+\bar{\eta}^{2}}\approx
-(0.02\pm 0.006)$ [4], $\phi _{M}=-\beta ;$ for $B_{s}^{0},$ $r=-\sqrt{\bar{%
\rho}^{2}+\bar{\eta}^{2}}\approx -(0.40\pm 0.13)$ [4], $\phi _{M}=0$. First
term of Eq.(\ref{45}) has an important implication: This term is zero, if $%
\alpha _{f}=-\bar{\alpha}_{\bar{f}};$ $\bar{\alpha}_{f}^{\prime }=-\alpha _{%
\bar{f}}^{\prime }$ as implied by $CP$-conservation. The finite value of
this term would imply $CP$ violation in baryon decay. The above equations
are simplified if we assume the validity of Eq.(\ref{24}). In that case we
have 
\begin{eqnarray}
\mathcal{A(}t) &=&\frac{2r\sin \Delta mt\sin (2\phi _{M}-\gamma )}{1+r^{2}}
\label{46} \\
\Delta \mathcal{A(}t) &=&0  \label{47} \\
\mathcal{F}(t) &=&\frac{1-r^{2}}{1+r^{2}}\cos \Delta mt  \label{48} \\
\Delta \mathcal{F}(t) &=&\frac{1-r^{2}}{2(1+r^{2})}(\alpha _{f}+\bar{\alpha}%
_{\bar{f}})\cos \Delta mt  \nonumber \\
&&-\frac{4r\sin \Delta mt\sin (2\phi _{M}-\gamma )S_{f}P_{f}}{%
(1+r^{2})(S_{f}^{2}+P_{f}^{2})}  \label{49}
\end{eqnarray}
Eq.(\ref{46}) gives a means to determine the weak phase $2\beta +\gamma $ or 
$\gamma $ in the baryon decay modes of $B_{d}^{0}$ and $B_{s}^{0}$
respectively. Non-zero $\cos \Delta mt$ term in $\Delta \mathcal{F}(t)$
would give clear indication of $CP$ violation especially for baryon decay
modes of $B_{d}^{0},$ for which \thinspace $r^{2}\leq 1,$ so that $\frac{%
1-r^{2}}{1+r^{2}}\approx 1$. Assuming $CP-$invariance, we get from Eqs.(\ref
{46}) and (\ref{49}) 
\begin{eqnarray}
-2S_{f}P_{f} &=&(S_{f}^{2}+P_{f}^{2})\frac{\Delta \mathcal{F}(t)}{\mathcal{A(%
}t)}  \nonumber \\
&=&\{(1.04\pm 0.38)\times 10^{10}\text{MeV}^{4}]\frac{\Delta \mathcal{F}(t)}{%
\mathcal{A(}t)}  \label{50}
\end{eqnarray}
The $S_{f}P_{f}$ can be determined from the experimental value of $\frac{%
\Delta \mathcal{F}(t)}{\mathcal{A(}t)}$ in future experiments.

The baryon decay modes of $B$-mesons not only provide a means to test
prediction of $CP$ asymmetry viz $\alpha _{f}+\bar{\alpha}_{\bar{f}}=0$ for
charmed baryons (discussed above) but also to test the $CP$-asymmetry in
hyperon (antihyperon) decays viz absence of $CP$-odd observables $\Delta
\Gamma ,\Delta \alpha ,\Delta \beta $ discussed in [3]. Consider for example
the decays 
\[
B_{q}^{0}\rightarrow p\bar{\Lambda}_{c}^{-}\rightarrow p\bar{p}K^{0}(p\bar{%
\Lambda}\pi ^{-}\rightarrow p\overline{p}\pi ^{+}\pi ^{-}),
\]
\[
\bar{B}_{q}^{0}\rightarrow \overline{p}\Lambda _{c}^{+}\rightarrow \bar{p}p%
\bar{K}^{0}(\overline{p}\Lambda \pi ^{+}\rightarrow \bar{p}p\overline{b}\pi
^{-}\pi ^{+})
\]
By analyzing the fianl state $\bar{p}p\bar{K}^{0},p\bar{p}K^{0},$ one may
test $\alpha _{f}=-\bar{\alpha}_{\bar{f}}$ for the charmed hyperon. We note
that for $\Lambda _{c}^{+},$ $c\tau =59.9\mu $m, whereas $c\tau =7.8$cm for $%
\Lambda -$hyperon [4], so that the decays of $\Lambda _{c}^{+}$ and $\Lambda 
$ would not interfere with each other. By analysing the final state $\bar{p}%
p\pi ^{-}\pi ^{+}$ and $p\bar{p}\pi ^{+}\pi ^{-},$ one may check $CP$%
-violation for hyperon decays. One may also note that for $(B_{d}^{0},\bar{B}%
_{d}^{0})$ complex, the competing channels viz $B_{d}^{0}\rightarrow \bar{p}%
\Lambda _{c}^{+},$ $\bar{B}_{d}^{0}\rightarrow p\bar{\Lambda}_{c}^{-}$ are
doubly Cabibbo supressed by $r^{2}=\lambda ^{4}\left( \bar{\rho}^{2}+\bar{%
\eta}^{2}\right) $ unlike $(B_{s}^{0}-\bar{B}_{s}^{0})$ complex where the
competing channels are supressed by a factor of $\left( \bar{\rho}^{2}+\bar{%
\eta}^{2}\right) $. Hence $B_{d}^{0}($ $\bar{B}_{d}^{0})$ decays are more
suitable for this type of analysis. Other decays of intrest are 
\begin{eqnarray*}
B^{-} &\rightarrow &\Lambda \bar{\Lambda}_{c}^{-}\rightarrow \Lambda \bar{%
\Lambda}\pi ^{-}\rightarrow p\pi ^{-}\bar{p}\pi ^{+}\pi ^{-} \\
B^{+} &\rightarrow &\bar{\Lambda}\Lambda _{c}^{+}\rightarrow \bar{\Lambda}%
\Lambda \pi ^{+}\rightarrow \bar{p}\pi ^{+}p\pi ^{-}\pi ^{+} \\
B_{c}^{-} &\rightarrow &\bar{p}\Lambda \rightarrow \bar{p}p\pi ^{-} \\
B_{c}^{+} &\rightarrow &p\bar{\Lambda}\rightarrow p\bar{p}\pi ^{+}
\end{eqnarray*}
The non-leptonic hyperon (antihyperon) decays $N\rightarrow N^{\prime }\pi (%
\bar{N}\rightarrow \bar{N}^{\prime }\bar{\pi})$ are related to each other by 
$CPT$%
\begin{eqnarray*}
a_{l}(I) &=&\left\langle f_{lI}^{out}\left| H_{W}\right| N\right\rangle
=\eta _{f}e^{2i\delta _{l}(I)}\left\langle \bar{f}_{lI}^{out}\left|
H_{W}\right| \bar{N}\right\rangle  \\
&=&\eta _{f}e^{2i\delta _{l}(I)}\bar{a}_{l}^{*}(I)
\end{eqnarray*}
Hence 
\[
\bar{a}_{l}(I)=\eta _{f}e^{2i\delta _{l}(I)}\bar{a}%
_{l}^{*}(I)=(-1)^{l+1}e^{i\delta _{l}(I)}e^{-i\phi }\left| a_{l}\right| 
\]
where we selected the phase $\eta _{f}=(-1)^{l+1}$. Here $I$ is the isospin
of the final state and $\phi $ is the weak phase. Thus necessary condition
for non-zero $CP$ odd observables is that the weak phase for each partial
wave amplitude should be different [see ref \cite{3} for details; for a
review see first ref in \cite{1}].

\textbf{Acknowledgments}

The author acknowledges a research grant provided by the Higher Education
Commission of Pakistan to him as a Distinguished National Professor.

\textbf{Figure Captions}

Figure1a: $W$-exchange diagram for $B_{q}^{0}\rightarrow N_{1}\bar{N}%
_{2}\left( M_{f}\right) $

Figure1b: $W$-exchange diagram for $\bar{B}_{q}^{0}\rightarrow N_{1}\bar{N}%
_{2}\left( \bar{M}_{f}^{\prime }\right) $

Figure2: Annihilation diagram for $B^{-}\rightarrow N_{1}\bar{N}_{2}$

\end{document}